# Dielectric Anomaly in Ice near 20 K; Evidence of Macroscopic Quantum Phenomena

Fei Yen,[1,2,*] Tian Gao,[3]

[1]Key Laboratory of Materials Physics, Institute of Solid State Physics, Hefei Institutes of Physical Science, Chinese Academy of Sciences, Hefei 230031, P. R. China
[2]High Magnetic Field Laboratory, Hefei Institutes of Physical Science, Chinese Academy of Sciences, Hefei 230031, P. R. China
[3]School of Mathematics and Physics, Shanghai University of Electric Power, Shanghai 200090, P. R. China

Correspondence:
[*]fyen18@hotmail.com or fyen@issp.ac.cn, +86-1334-929-0010,

**Abstract:** H₂O is one of the most important substances needed in sustaining life; but yet not much is known about its ground state. Here, a previously unidentified anomaly is identified in the form of a minimum in the imaginary part of the dielectric constant with respect to temperature near 20 K while the real part remains monotonic. Isothermal dispersion and absorption measurements show coinciding results. For the case of heavy ice (D₂O), no anomaly was identified confirming an apparent isotope effect. Concerted quantum tunneling of protons is believed to be the main cause behind the reported anomaly. Our findings identify another system that exhibits macroscopic quantum phenomena of which rarely occur in nature.

**TOC graphic:**

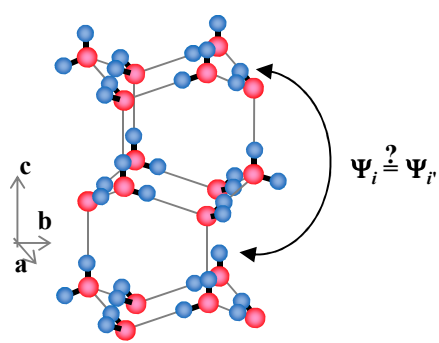

When water freezes, the oxygen atoms crystallize into a puckered hexagonal-like pattern [1]. According to the 'ice rules' [2], a) each oxygen atom has four hydrogen atoms linked to it: two via covalent bonding which make up the $H_2O$ molecule and two weaker intermolecular hydrogen bonds. b) Each hydrogen atom on the other hand, is linked to only two oxygen atoms: one by covalent bonding and one by hydrogen bonding. Since the lengths of the covalent and hydrogen bonds are different, between each lone pair of oxygen, two sites are available for the proton of the hydrogen atom to reside in (Fig. 1a). The above description is usually what is known as the ice I*h* phase of $H_2O$. At about $T_g$=136 K, the protons (as massive as they are compared to electrons) become prohibitively immobile [3] so at temperatures below $T_g$, the protons "freeze" into a glassy state while the oxygen atoms remain in the same hexagonal lattice. Since entropy is directly associated to disorder, the randomly frozen protons causes a residual entropy to be present at absolute zero [4]. Upon further cooling to 60 K [5], the hexagonal lattice of oxygen atoms undergoes a structural phase transition into the orthorhombic structure (space group *Cmc*$2_1$) [6] also known as the ice XI phase and presumed to be proton ordered. During warming, the phase transition back to Ice I*h* from ice XI occurs near 72 K [5,7,8]. Experiments at even lower temperature have proven to be complicated as in most cases some type of catalytic stimulant is needed to accelerate the formation of ice XI domains [8,9]. In a previous work, we reported a new method of obtaining *pure* $H_2O$ ice XI domains by employing $H_2O$ itself as a pressure medium with embedded platinum electrodes and pressurizing only in the liquid state [5].

In this Letter, we extend the study of the dielectric properties of undoped $H_2O$ ice down to 5 K under different hydrostatic pressures and find that an anomaly in the form of a minimum in the imaginary part of the dielectric constant exists near 20 K. The cause behind the anomaly is believed to stem from a concerted quantum tunneling of protons across their adjacent empty sites. Our results classify $H_2O$ as another system that exhibits macroscopic quantum phenomena, of which is only the second system that is based on fermions, the other one being superconductivity.

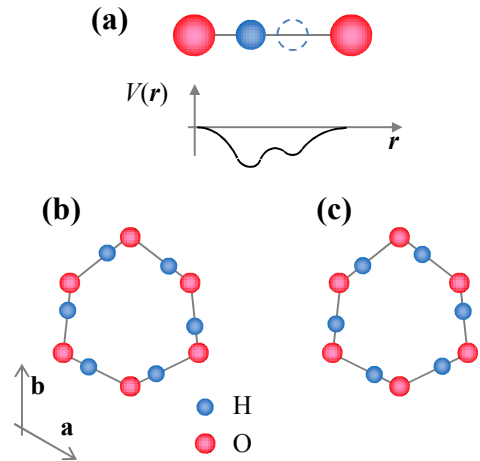

**Figure 1.** (a) Each proton lying in between a pair of oxygen atoms experiences a double-well type potential $V(r)$ due to the covalent and hydrogen bonds. At low enough temperature, a proton can tunnel through to its adjacent site (dashed blue circle). However, tunnelling of one single proton would violate the ice rules and result in an oxygen atom to be covalently bonded to three protons and another oxygen atom to only one proton. If six protons within a hexamer ring tunnel at the same time such as from configuration (b) to (c), then the ice rules are preserved. Note how the dipole moment distribution of (b) and (c) are both zero suggesting that such a process should not induce a change in the real part of the dielectric constant.

Figure 2 shows the cooling and warming curves of the imaginary part of the dielectric constant $\varepsilon''(T)$ with respect to temperature at the constant hydrostatic pressure of 33 MPa. The newly identified feature in this curve is the minimum at $T_Q$=20 K. $T_{Ih\text{-}XI}$ and $T_{XI\text{-}Ih}$ are the onset temperatures of the phase transitions from ice $Ih$ to XI and vice versa, respectively. $T_{PO}$=100 K is the temperature at which continuous proton ordering takes place [5]. The inset of Fig. 2 shows an enlarged view of the cooling and warming curves of $\varepsilon''(T)$ at low temperature where a hysteretic region of 2 K associated to $T_Q$ was observed. At $T<T_Q=20$ K, the upswing in $\varepsilon''(T)$ is believed to stem from a correlated quantum tunneling of the protons as will be discussed below.

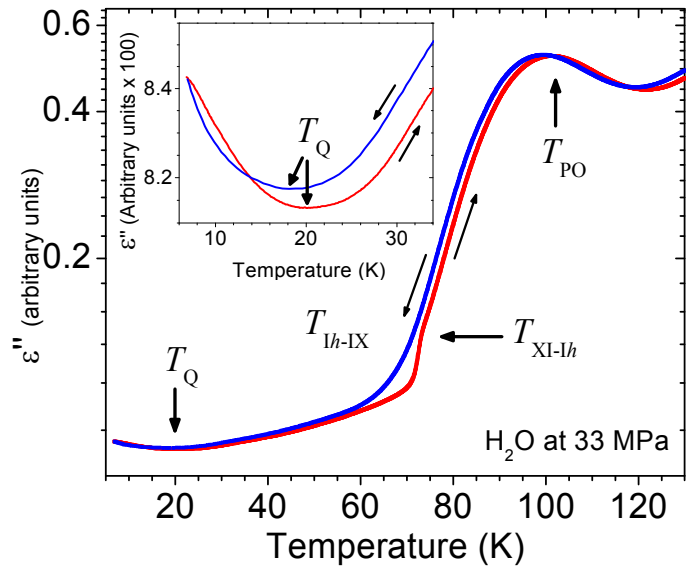

**Figure 2.** Imaginary part of the dielectric constant $\varepsilon''(T)$ with respect to temperature showing the onset of proton ordering and disordering at $T_{PO}=100$ K and the transitions into and out of ice XI from ice I$h$ at $T_{Ih\text{-}XI}=59$ K and $T_{XI\text{-}Ih}=73$ K, respectively. Inset is an enlarged view of the low temperature region showing that a minimum exists in $\varepsilon''(T)$ near $T_Q=20$ K with a hysteresis of ~2 K. Correlated tunneling of the protons is believed to take place below $T_Q$.

Figure 3 shows the warming curves of $\varepsilon''(T)$ at various hydrostatic pressures down to 7 K. The inset shows the real part of the dielectric constant $\varepsilon'(T)$ which did not exhibit any anomalies near $T_Q$; this is consistent with thermal expansion and lattice constant ratio results where monotonic behavior was reported to occur in the same temperature range all the way down to about 10 K [10]. In addition, a curve for the case of heavy ice ($D_2O$) is also included where a minimum was not observed in both $\varepsilon'(T)$ and $\varepsilon''(T)$ confirming an apparent isotope effect [11].

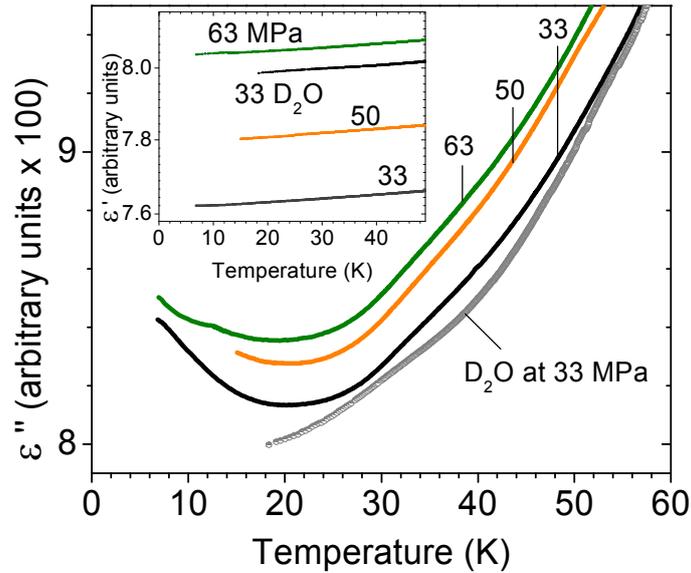

**Figure 3.** Warming curves of ε"(*T*) at different hydrostatic pressure. The D$_2$O curve is offset by 0.35 for clarity. Inset: real part of the dielectric constant ε'(*T*) with respect to temperature.

Figures 4a and 4b show, respectively, the dispersion ε'(ω) and absorption ε"(ω) curves of H$_2$O at 150 MPa for different temperature. With decreasing temperature, ε'(ω) decreased systematically throughout the studied frequency range. For ε"(ω) on the other hand, the absorption curve at 16 K was the lowest in magnitude which is consistent with the observation of a minimum in ε"(*T*) near 20 K (Figs. 2 and 3). Fig. 4c shows ε'(ω) and ε"(ω) plotted as the abscissa and ordinate (matched to their frequencies) for different temperatures, respectively. Usually, a semicircle is formed for the case of polar dielectrics which stems from a classical treatment of the molecules along with the addition of a damping term [12]. The increase in curvature of the 5 and 10 K semicircles clearly indicate that a classical approach is not sufficient and that some type of quantum phenomena takes place at such low temperatures.

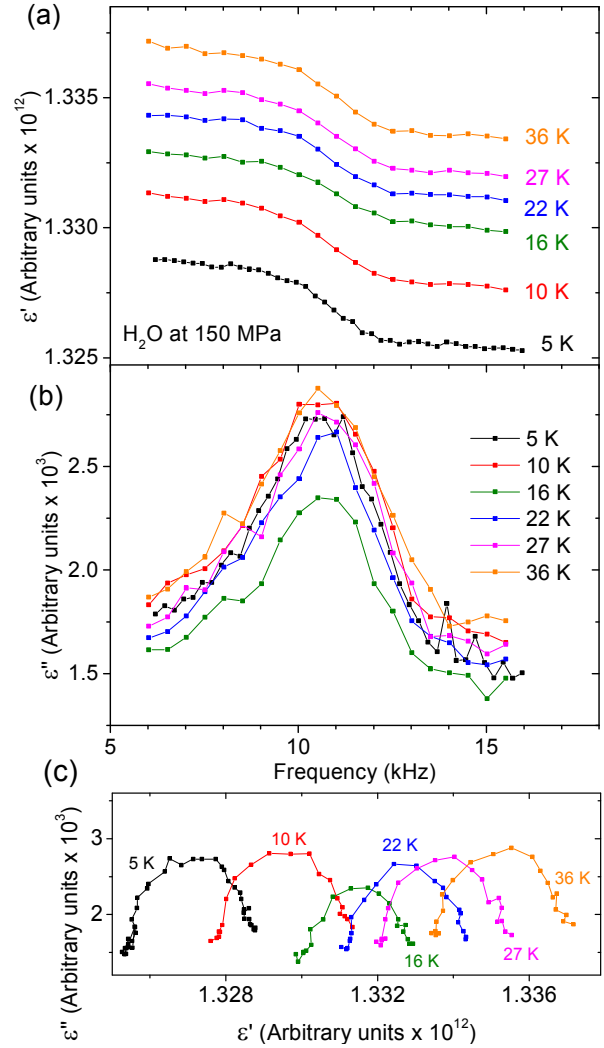

**Figure 4.** (a) Dispersion ε'(ω) and (b) absorption ε"(ω) data of $H_2O$ at different isotherms from 5 to 36 K. (c) Cole-Cole representations at different temperature. The deviation from the semicircular shape of the 5 and 10 K curves suggests an emergence of quantum phenomena at low temperature.

Here is a case where an anomalous change occurs in ε"(*T*) in the form of a minimum while ε'(*T*) remains monotonic in the same temperature range. Even though the oxygen atoms lie in a hexagonal lattice, they are calculated to be about 0.008 Å off-centered from their sites at 0 K [11]. Any physical displacement of the equilibria of the oxygen or hydrogen atoms should be reflected as an anomaly in ε' so the possibility of a further ordering of the off-centered oxygen atoms near 20 K can be ruled out. The physical meaning of ε" is the amount of energy dissipated when an external electric field is switched so a minimum in ε"(*T*) while ε'(*T*) remains monotonic can be interpreted as an increase in the rotation and/or reorientation of the water molecules. However, the fundamental frequencies of the intra-molecular, translation and rotation vibrations in ice are extremely high [13] while the minimum in ε"(*T*) was observed at frequencies lower than 16 kHz.

Many phenomenological models are very accurate in describing the properties of water and ice but all of them have their own limitations [14]. Most of these models only take into consideration the classical aspects of the system whereas the inter-molecular bonds in solid $H_2O$ have been shown to otherwise be quantum in nature through Compton scattering experiments [15]. Other recent incorporations of quantum effects to existing theoretical models have yielded extremely accurate explanations of some of the anomalous behavior in ice such as its unusual isotope effect near 0 K [11]; density minimum near 60 K [16]; and heat capacity near 0 K [17]. Moreover, Hernandez de la Peña *et al.* have shown that lattice vibrations are indeed affected by quantization [18], though their work was only constrained to 160–235 K. Visualization of up to four protons tunneling in concert at 5 K was recently reported [19]. Castro Neto *et al.* concluded that quantum tunneling of protons to their adjacent empty site is what makes the ice I*h*/XI phase transition possible since all classical movement of the protons near 72 K is suppressed [20]. Bove *et al*. reported observing jump distances of 75 Å [21] which is nearly equal to the distance between the two available proton sites in between each pair of oxygen atoms (Fig. 1a). According to the 'ice rules', if a single proton undergoes tunneling, then an OH- site is created and the ice rules are violated. However, if six protons within a hexamer ring were to tunnel at the same time such as from the configuration shown in Fig. 1b to Fig. 1c, then the ice rules are preserved. In fact, recent quantum mechanical *ab initio* path integral simulations have shown that concerted tunneling of six protons in a hexamer ring is possible at low temperatures and ambient pressure [22]. Such a process should not be reflected in $\varepsilon'(T)$ because the initial (Fig. 1b) and final (Fig. 1c) equivalent dipole moment distributions are both zero. However, $\varepsilon''(T)$ should exhibit an increase because there is a movement of charges. Hence, the anomalous behavior observed in $\varepsilon''(T)$ in our data appears to strictly stem from the back and forth tunneling of protons in groups of six. This interpretation is also consistent with a recent study suggesting that the protons possibly behave as a quantum liquid fluctuating continually at low temperatures [23].

Moreover, the absence of a minimum in $\varepsilon''(T)$ in $D_2O$ indicates that concerted proton tunneling does not take place in heavy ice which is consistent with recent reports where: a) a 20% substitution of deuterium reportedly hindered the observed proton jumps in pure $H_2O$ [21]; b) the probability of deuterons tunneling in concert at 5 K is expected to be over two orders of magnitude lower than protons [19]; and c) partial to full deuteration of ice inhibits collective proton tunneling [24].

For several protons to tunnel in concert, their wave functions must be the same so they must all occupy the same state; and in this case most likely the ground state. The same can be argued for a large number of protons such as the few percent of protons undergoing concerted tunneling within a crystal observed experimentally by Bove *et*

*al*. [21]. However, according to the Pauli Exclusion Principle, protons have spin 1/2 so they cannot all simultaneously occupy the ground state. Hence, we conjecture that all tunneling of the protons must occur in pairs so that a quasiparticle of spin 1, such as the case of a Cooper pair being formed by two electrons mediated by a phonon in superconductors [25], would allow for the paired protons to also act as a boson and collapse into the ground state to form a Bose condensate. The observed increase in $\varepsilon''(T)$ with decreasing temperature suggests that more and more protons become involved in the correlated tunneling process which lends evidence to the occurrence of a new type of macroscopic quantum phenomena. More work is definitely needed on developing a microscopic theory for this new phenomenon of correlated proton tunneling.

The authors wish to thank Prof. J. L. F. Abascal for insightful suggestions during the early stages of this work. This work was made possible by the National Natural Science Foundation of China, grant numbers 11374307 and 11204171.

**Methods:**

The sample was prepared using de-ionized, 18 MΩ $H_2O$ from Milli-Q poured inside a Teflon container and secured inside a double piston clamp cell made of BeCu. Specific details regarding the procedure are found in Ref [5]. Once the pressure cells have been pressurized and locked into position at room temperature, they are mounted onto a probe and inserted into a cryostat. The real and imaginary parts of the dielectric constant were obtained from measuring the capacitance and loss tangent, respectively, of a pair of platinum plates positioned 0.5 mm apart inside the Teflon container.

*Temperature dependent measurements*. A customized cryostat made by Janis Corp was employed. The cooling and warming rates were ~2 K/min. An Andeen Hagerleen AH2700A capacitance bridge operating at 1 kHz was employed to simultaneously measure the real and imaginary parts of the dielectric constant. The diameter of the pistons of the pressure cell was 5 mm.

*Dispersion and absorption measurements*. An Agilent E4980A LCR Meter was employed to measure the real and imaginary parts of the dielectric constant. Temperature was controlled by a commercial cryostat (Dynacool Physical Properties Measurement System by Quantum Design). The sample was first cooled from room temperature down to 5 K at 2 K/min. For each temperature scan, it was made sure that the temperature remained stable for 5 minutes. A different pressure cell with a piston diameter of 4 mm was employed for these experiments.